\documentclass[journal]{IEEEtran}
\usepackage{amsmath,amsfonts}
\usepackage{algorithmic}
\usepackage{algorithm}
\usepackage{array}
\usepackage[caption=false,font=normalsize,labelfont=sf,textfont=sf]{subfig}
\usepackage{textcomp}
\usepackage{stfloats}
\usepackage{url}
\usepackage{verbatim}
\usepackage{graphicx}
\usepackage{cite}
\usepackage[hidelinks]{hyperref}   
\usepackage[nameinlink,noabbrev]{cleveref} 

\usepackage{xcolor}
\usepackage{amsmath} 
\usepackage{multirow} 
\usepackage[table]{xcolor}
\definecolor{darkmossgreen}{RGB}{180, 210, 180}
\definecolor{dashedarrow}{HTML}{00B2B2}
\usepackage{booktabs}
\usepackage{makecell}
\usepackage{tabularx}
\usepackage{adjustbox}
\usepackage{ragged2e}
\usepackage{enumitem}
\usepackage{threeparttable}

\begin{document}

\title{RDDM: A Residual-Driven Drifting Model for High-Fidelity Low-Dose CT Denoising}

\author{Jianxu Wang, Qing Lyu, Ge Wang,~\IEEEmembership{Fellow,~IEEE}
\thanks{Jianxu Wang and Ge Wang are with the Biomedical Imaging Center, Department of Biomedical Engineering, Rensselaer Polytechnic Institute, Troy, NY 12180 USA (e-mail: wangj68@rpi.edu; wangg6@rpi.edu).}
\thanks{Qing Lyu is with the Department of Radiology and Biomedical Imaging, Yale School of Medicine, New Haven, CT 06510 USA (e-mail: qing.lyu@yale.edu).}
\thanks{Code is available at: \url{https://github.com/Jayx-Wang/RDDM}.}
}

\maketitle

\begin{abstract}
Low-dose CT (LDCT) denoising remains an important yet challenging problem in medical imaging. Although recent learning-based methods have shown promising performance, those optimized using classical pixel-level objectives often produce over-smoothed reconstructions. Existing mainstream generative models, such as diffusion models, have improved fidelity at the cost of expensive multi-step iterative inference, which limits their practicality for real-time use.
To address this gap, we propose a Residual-Driven Drifting Model (RDDM) for effective, efficient, and high-fidelity LDCT denoising. Inspired by the recently proposed Drifting Models, RDDM incorporates the multi-step distribution evolution into the training dynamics through a residual drifting field, thereby enabling one-step denoising. Specifically, the residual drifting field is formed by an attractive force induced by the residuals between LDCT and normal-dose CT (NDCT) and a repulsive force induced by the generated residuals. In addition, by adjusting the parameter settings and incorporating pixel-level supervision, we develop three RDDM variants, covering application needs from detail preservation to stronger noise suppression.
Extensive experiments demonstrate that RDDM achieves state-of-the-art denoising performance among supervised baselines. In particular, RDDM-Fine produces reconstructions that are highly consistent with NDCT, achieving superior PSNR and SSIM together with the best FID of 5.87 while preserving realistic anatomical textures. Moreover, RDDM enables on-the-fly inference, requiring only about 15 ms to denoise a single $512\times512$ LDCT slice. These results establish RDDM as a promising solution for high-fidelity and real-time LDCT denoising in clinical applications.
\end{abstract}


\section{Introduction}
\IEEEPARstart{C}{omputed} tomography (CT) is a widely used medical imaging modality in modern clinical practice, providing rapid cross-sectional imaging of internal anatomical structures~\cite{smith2025projected}. It plays an indispensable role in disease screening, diagnosis, and treatment planning for a broad range of clinical conditions. The reliability of these clinical applications strongly depends on high-quality CT images with clear anatomical boundaries and faithful tissue textures. However, CT imaging is inevitably accompanied by ionizing radiation. The radiation exposure associated with normal-dose CT (NDCT) may pose potential health risks to patients, especially for pediatric populations and individuals requiring repeated examinations~\cite{brenner2007computed}. Consequently, achieving low radiation dose while preserving high diagnostic image quality has become an important objective in CT imaging.

A common strategy for radiation dose reduction is low-dose CT (LDCT) acquisition. However, lowering the X-ray photon flux leads to reduced measurement statistics, resulting in increased quantum noise and reduced visibility of subtle anatomical structures. These degradations may obscure subtle lesions, distort anatomical details, and thereby compromise diagnostic accuracy. Therefore, image denoising techniques that can recover high-quality images from LDCT scans while preserving clinically relevant structures and textures are in urgent demand.

Traditional denoising methods generally rely on hand-crafted priors about image structures and noise properties~\cite{lindenbaum1994gabor,moulin1999analysis,dabov2007image,sidky2008image,manduca2009projection}. Based on these priors, noise suppression is performed through techniques such as filtering, iterative correction, and regularized optimization. However, such methods generally perform well only when the assumed noise model matches the actual noise distribution. In addition, they often require manual tuning of hyperparameters to accommodate different noise levels and imaging conditions. For example, BM3D requires an estimated noise standard deviation ($\sigma$) as an input parameter for effective denoising, which often needs to be determined through repeated trial-and-error based on the noise level. More importantly, aggressive filtering tends to suppress subtle anatomical textures together with noise, leading to loss of fine details and over-smoothed reconstructions.

Over the past decade, rapid advances in deep learning theory and computational hardware have led to remarkable progress in image denoising~\cite{tian2020deep,kulathilake2023review}. Compared with traditional model-based approaches, data-driven denoising methods have demonstrated strong capability in handling complex real-world noise and recovering high-quality image content.
Existing data-driven denoising methods are mainly trained under supervised and self-supervised (or unsupervised) paradigms. 
Supervised learning typically relies on paired training data. In particular, methods trained with clean--noisy image pairs are commonly referred to as Noise2Clean approaches~\cite{chen2017low,shan2019competitive,wang2023ctformer}. To relax the requirement for clean targets, Noise2Noise~\cite{lehtinen2018noise2noise} was proposed to enable training with noisy--noisy image pairs. However, such paired noisy observations must still originate from the same clean image and be corrupted by independent zero-mean noise. Moreover, its performance approaches that of Noise2Clean only in the large-data regime under ideal noise assumptions. Nevertheless, many CT denoising methods~\cite{fang2021iterative,wu2021low,wu2022deep} built upon the Noise2Noise framework have achieved promising results.
Self-supervised learning further reduces the dependence on paired data or manual annotations, but its denoising performance is generally inferior to that of supervised learning~\cite{krull2019noise2void,batson2019noise2self,quan2020self2self,niu2022noise}. More recently, zero-shot learning under the self-supervised paradigm has gone a step further by removing the dependence on a training dataset altogether, requiring only a single noisy image for training and direct denoising. Recent studies~\cite{lequyer2022fast,wang2025zs,mansour2023zero,wang2025median2median} have shown that zero-shot methods can achieve performance competitive with dataset-driven approaches in certain scenarios.

Despite the differences in training paradigms, most of the aforementioned deep learning-based denoising methods are optimized directly with classical pixel-wise losses such as $\ell_1$ or $\ell_2$, which often favor smoother reconstructions. For CT imaging, such over-smoothing may suppress important anatomical details together with noise, leading to loss of structural and textural fidelity in the denoised images.

\begin{figure*}[t]
\centering
\includegraphics[width=\linewidth]{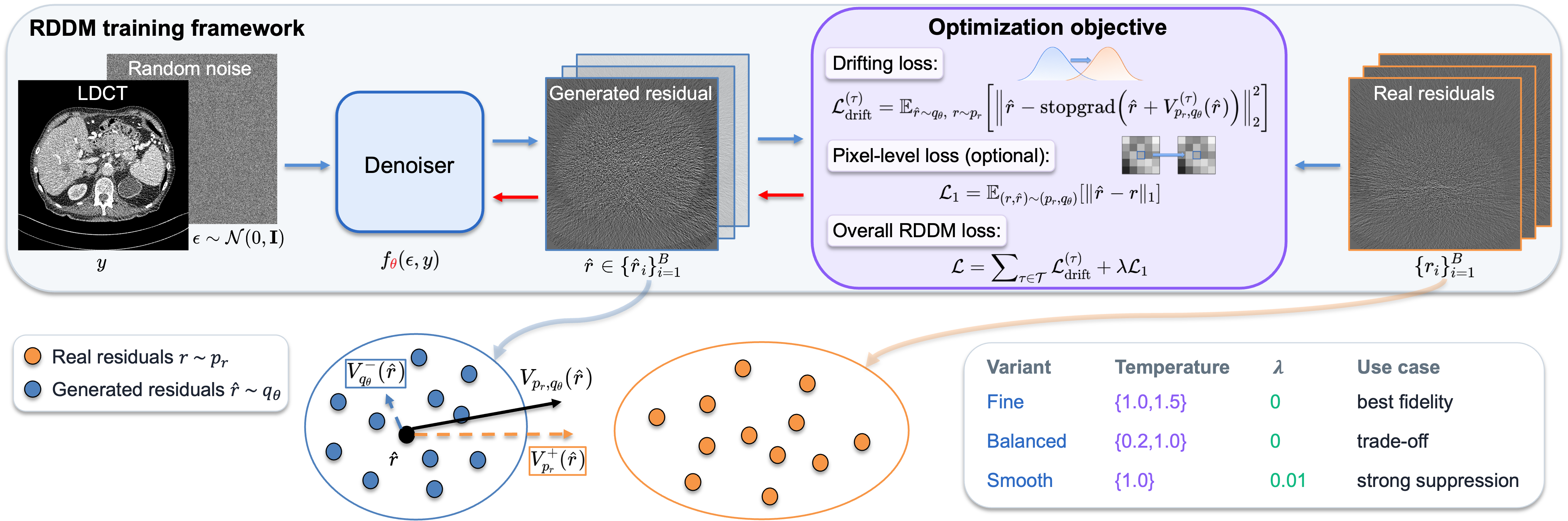}
\caption{Overview of the proposed RDDM framework. Top: the RDDM training framework, where a single residual sample is generated by the denoiser, while the optimization objective is computed using the full generated residual batch and real residual batch of size $B$. Bottom left: illustration of drifting a generated sample toward the real residual distribution during training-time evolution. Bottom right: summary of the three RDDM variants under different settings.}
\label{fig:overview}
\end{figure*}

To better preserve realistic image textures and improve reconstruction fidelity, recent studies~\cite{ho2020denoising,song2020denoising,xu2022poisson,xu2023pfgm++} have increasingly explored generative models with strong distribution modeling capacity. 
In particular, denoising diffusion probabilistic models (DDPMs)~\cite{ho2020denoising} have attracted widespread attention due to their ability to model complex data distributions. 
DDPMs define a forward diffusion process that gradually perturbs training images with noise, and learn a reverse denoising process that progressively maps random noise back to the target data distribution. This multi-step iterative formulation enables the model to generate high-fidelity reconstructions. Denoising diffusion implicit models (DDIMs)~\cite{song2020denoising} further improve the sampling efficiency of DDPMs by introducing a non-Markovian reverse process, allowing high-quality reconstructions to be generated with fewer sampling steps. Diffusion-based methods have shown promising performance in LDCT denoising~\cite{gao2023corediff,gao2022cocodiff}, particularly in recovering rich texture details and producing image appearances similar to NDCT. 
More recently, flow-based generative models, including the Poisson flow generative model (PFGM)~\cite{xu2022poisson} and its generalized version PFGM++~\cite{xu2023pfgm++}, have emerged as promising alternatives to diffusion models, with favorable stability and sampling efficiency. By applying PFGM++ to low-dose CT denoising, PPFM~\cite{hein2024ppfm} and FORCE~\cite{xia2026tomographic} have further demonstrated strong denoising performance with high data fidelity.

However, both diffusion models and flow-based models rely on multiple iterative sampling steps during inference to achieve high-quality reconstructions, resulting in substantial computational cost for denoising. Although some methods~\cite{song2020denoising,hein2024ppfm,xia2026tomographic} attempt to reduce the number of function evaluations (NFEs), this usually comes at the expense of reconstruction quality, while still requiring dozens of NFEs to reach stable convergence. Consequently, these methods remain less suitable for real-time clinical applications. Therefore, there is still a need for an LDCT denoising method for clinical diagnosis that can simultaneously provide highly faithful reconstructions and achieve both effective and efficient denoising.

Recently, Drifting Models~\cite{deng2026drifting} were proposed as a new generative modeling framework that reformulates the multi-step distribution evolution in existing mainstream generative models~\cite{ho2020denoising,xu2022poisson,xu2023pfgm++} into a training-time evolution driven by a drifting field. Inspired by this idea, we propose a Residual-Driven Drifting Model (RDDM) for effective and efficient low-dose CT denoising. 

Fig.~\ref{fig:overview} provides an overview of the proposed RDDM framework. In this framework, the denoiser takes a random noise $\epsilon \sim \mathcal{N}(0,\mathbf{I})$ as input and an LDCT image $y$ as the condition to generate the corresponding residual estimate $\hat r$. In each training iteration, a batch of generated residuals $\{\hat r_i\}_{i=1}^{B}$ is produced as an empirical approximation of the generated residual distribution $q_\theta$. This generated residual batch, together with the real residual batch $\{r_i\}_{i=1}^{B}$ computed from LDCT--NDCT pairs, is then used to optimize the denoiser. The optimization objective consists of a distribution-level drifting loss and an optional pixel-level reconstruction loss. In the residual space, each generated residual $\hat r$ is driven by a residual drifting field $V_{p_r,q_\theta}(\hat r)$, where generated residuals contribute repulsion and real residuals contribute attraction. As training proceeds, this residual drifting field pushes the generated residual distribution toward the real residual distribution. Building upon this framework, we further develop three RDDM variants, namely RDDM-Fine, RDDM-Balanced, and RDDM-Smooth, through different temperature settings and pixel-level reconstruction supervision to accommodate different application requirements. The corresponding parameter settings are summarized in the bottom-right panel of Fig.~\ref{fig:overview}.

Experimental results show that RDDM-Smooth achieves PSNR and SSIM comparable to conventional $\ell_1$-supervised methods while obtaining better FID, whereas RDDM-Fine not only surpasses state-of-the-art supervised baselines in PSNR and SSIM, but also achieves the best FID. These results indicate that the reconstructions produced by RDDM-Fine are highly consistent with NDCT at both the pixel level and the distribution level, demonstrating strong fidelity and realistic texture preservation.
Moreover, RDDM requires only 1 NFE for denoising and takes about 15 ms to process a single $512\times512$ LDCT slice, enabling on-the-fly deployment. The combination of high-fidelity reconstruction and fast denoising gives RDDM strong potential for real-time and clinical applications.

\section{Methodology}
\subsection{Drifting Models Recap}
\label{sec:drifting_models}
Drifting Models~\cite{deng2026drifting} leverage the iterative optimization of network parameters to absorb the evolution of the pushforward distribution into the training process. Its central idea is to learn a one-step generator whose outputs are progressively corrected by a drifting field constructed from target-guided attraction and model-aware repulsion. 

Let $\epsilon$ be a random variable sampled from a base distribution $p_0$, and let $f_{\theta}$ denote a neural generator parameterized by $\theta$. The generated sample is given by $x = f_{\theta}(\epsilon) \sim q_{\theta}$. The induced model distribution is the pushforward of $p_0$ under $f_{\theta}$, written as
\begin{equation}
q_{\theta} = (f_{\theta})_{\#} p_0,
\end{equation}
where the goal of generative modeling is to make the model distribution $q_{\theta}$ approach the target data distribution $p$.

Traditional generative models, such as diffusion~\cite{sohl2015deep,song2020score} and some flow-based models~\cite{lipman2022flow,liu2022flow}, progressively refine generated samples via explicit iterative updates of the form $x_{i+1} = x_i + \Delta x_i$, where $\Delta x_i$ is estimated by a generator $f_{\theta}$ and $i$ indexes the inference iteration. Such a formulation requires multi-step inference. Drifting Models replace this explicit iterative refinement with the iterative nature of network optimization itself. Specifically, over $T$ training steps, a sequence of pushforward distributions $\{q_{\theta_i}\}_{i=1}^{T}$ is produced, where $q_{\theta_i}=(f_{\theta_i})_{\#} p_0$.
Accordingly, the implicit update for a generated sample $x_i$ at training iteration $i$ can be written as
\begin{equation}
x_{i+1} = x_i + \left(f_{\theta_{i+1}}(\epsilon) - f_{\theta_i}(\epsilon)\right)
= x_i + \Delta x_i,
\end{equation}
which indicates that each training iteration naturally induces an update of the generated distribution. As a result, inference can be performed in a single forward pass after training.

To formalize this idea, Drifting Models introduce a function $V_{p,q_\theta}$, referred to as a drifting field, which specifies how a generated sample $x$ should move under the interaction between the target distribution $p$ and the current model distribution $q_\theta$. It therefore estimates the displacement $\Delta x$ for $x$.
At training iteration $i$, the evolution of a generated sample follows
\begin{equation}
x_{i+1} = x_i + V_{p,q_i}(x_i),
\end{equation}
where $x_i = f_{\theta_i}(\epsilon) \sim q_{\theta_i}$, and the updated sample $x_{i+1}$ follows a new distribution $q_{\theta_{i+1}}$. Specifically, the drifting field is decomposed into an attraction term from the target distribution and a repulsion term from the model distribution. Given a sample $x\sim q_{\theta}$, the drifting field is defined as
\begin{equation}
V_{p,q_\theta}(x)=V_p^{+}(x)-V_{q_\theta}^{-}(x).
\end{equation}
The term $V_p^{+}(x)$ attracts the sample $x$ toward the target distribution $p$, while $V_{q_\theta}^{-}(x)$ repels the sample from the current model distribution $q_\theta$.

The drifting field defined in this way satisfies an anti-symmetric property: $V_{p,q}(x) = -V_{q,p}(x), \forall x$, which ensures that the drift vanishes when the model distribution matches the target distribution. Based on this equilibrium perspective, the optimal generator can be interpreted as satisfying the following fixed-point condition:
\begin{equation}
f_{\hat{\theta}}(\epsilon)
=
f_{\hat{\theta}}(\epsilon)
+
V_{p,q_{\hat{\theta}}}\!\left(f_{\hat{\theta}}(\epsilon)\right).
\end{equation}
This leads to a practical training objective in which the current prediction is regressed toward its drifted but frozen target:
\begin{equation}
\label{eq:loss}
\mathcal{L}_{\mathrm{drift}}
=
\mathbb{E}_{\epsilon \sim q_{0}}
\left[
\left\|
f_{\theta}(\epsilon)
-
\operatorname{stopgrad}
\!\left(
f_{\theta}(\epsilon)
+
V_{p,q_{\theta}}\!\left(f_{\theta}(\epsilon)\right)
\right)
\right\|_2^2
\right].
\end{equation}
Here, $\operatorname{stopgrad}(\cdot)$ prevents gradients from propagating through the drift target. 

\subsection{Residual Drifting for Generative Denoising}
We extend Drifting Models to a conditional formulation for image denoising by modeling the evolution of residual distributions. Assume that the degradation in LDCT images follows a latent noise distribution $p_r$, with paired NDCT images as reference. In practice, the noise distribution $p_r$ can be approximated by the residual distribution between NDCT and LDCT images. Given a set of LDCT images $Y$ and the corresponding NDCT images $X$, we define the residual set as $R$ with
\begin{equation}
r = y - x \in R, \quad r \sim p_r,
\end{equation}
where $(x, y)$ are paired samples with $x \in X$ and $y \in Y$, and $p_r$ serves as the target distribution in the denoising task.
Then, we introduce a conditional generator $f_{\theta}(\cdot, y)$ that maps a noise variable $\epsilon \sim \mathcal{N}(0,\mathbf{I})$ to a residual sample $\hat r$, i.e.,
\begin{equation}
\hat r = f_{\theta}(\epsilon, y)\sim q_\theta(r \mid y),
\end{equation}
where $y \in Y$ denotes an LDCT image that conditions the generation of $\hat r$, and $q_\theta(r \mid y)$ denotes the conditional distribution of the generated residual. Under this formulation, the objective of drifting-based denoising is to optimize the network parameters $\theta$ such that the conditional pushforward distribution $q_\theta(r \mid y)$ approaches the target residual distribution $p_r$, i.e.,
\begin{equation}
q_\theta(r \mid y) \approx p_r.
\end{equation}

Following the drifting formulation introduced in Sec.~\ref{sec:drifting_models}, we construct a drifting field in the residual space to guide the evolution of the generated residual distribution toward the target residual distribution $p_r$. 
Given a generated residual sample $\hat r \sim q_{\theta}(r \mid y)$, the drifting field is defined as
\begin{equation}
V_{p_r,q_\theta}(\hat r) = V_{p_r}^{+}(\hat r) - V_{q_\theta}^{-}(\hat r),
\end{equation}
where $V_{p_r}^{+}$ denotes the attraction toward the target residual distribution $p_r$, and $V_{q_\theta}^{-}$ denotes the repulsion from the current model distribution $q_\theta(r \mid y)$. 

Specifically, these components are estimated using mini-batch samples. Let $R=\{r_j\}_{j=1}^{B}$ denote the residual set and $\hat R=\{\hat r_i\}_{i=1}^{B}$ denote the generated residual set. 
The attraction component is estimated as
\begin{equation}
V_{p_r}^{+}(\hat r_i)
=
\frac{1}{Z_p(\hat r_i)}
\sum_{j=1}^{B}
k(\hat r_i,r_j)(r_j-\hat r_i),
\end{equation}
while the repulsion component is estimated as
\begin{equation}
V_{q_\theta}^{-}(\hat r_i)
=
\frac{1}{Z_{q_\theta}(\hat r_i)}
\sum_{j=1}^{B}
k(\hat r_i,\hat r_j)(\hat r_j-\hat r_i).
\end{equation}
Here $k(\cdot,\cdot)$ denotes a similarity kernel defined as
\begin{equation}
k(r, r') = \exp\left(-\frac{1}{\tau}\|r - r'\|\right),
\label{eq:similarity_kernel}
\end{equation}
where $\tau > 0$ is a temperature parameter that controls the bandwidth of the kernel. 
The normalization factors are defined as
\begin{equation}
Z_{p_r}(\hat r_i) = \sum_{j=1}^{B} k(\hat r_i, r_j), \quad
Z_{q_\theta}(\hat r_i) = \sum_{j=1}^{B} k(\hat r_i, \hat r_j).
\end{equation}

Accordingly, the residual drifting loss defined on a mini-batch is given by
\begin{equation}
\mathcal{L}_{\text{drift}}^{(\tau)}
=
\frac{1}{B} \sum_{i=1}^{B}
\left\|
\hat r_i
-
\operatorname{stopgrad}\!\left(
\begin{aligned}
\hat r_i
+\, V_{p_r,q_\theta}^{(\tau)}(\hat r_i)
\end{aligned}
\right)
\right\|_2^2,
\end{equation}
where $V_{p_r,q_\theta}^{(\tau)}(\hat r_i)$ denotes the drifting field at temperature $\tau$, and $\operatorname{stopgrad}(\cdot)$ denotes the stop-gradient operator.

In addition to the drifting loss $\mathcal{L}_{\text{drift}}$, we further incorporate a reconstruction loss $\ell_1$, defined as
\begin{equation}
\mathcal{L}_{1} = \frac{1}{B} \sum_{i=1}^{B} \| \hat{r}_i - r_i \|_1.
\end{equation}
The overall training objective is given by
\begin{equation}
\label{eq:loss_total}
\mathcal{L}
=
\sum_{\tau \in \mathcal{T}} \mathcal{L}_{\text{drift}}^{(\tau)}
+
\lambda \mathcal{L}_{1},
\end{equation}
where $\mathcal{T}$ denotes a set of temperature parameters for the similarity kernel, and $\lambda$ is a weighting parameter.

\subsection{Network and Training Details}
Our drifting-based denoising model adopts a U-Net architecture~\cite{ronneberger2015u} as the conditional generator $f_{\theta}$. A noise sample $\epsilon \sim \mathcal{N}(0,\mathbf{I})$ is concatenated with the LDCT image $y$ along the channel dimension and fed into $f_{\theta}$ to estimate the latent noise in $y$, i.e., $\hat r = f_{\theta}(\epsilon, y)$. The denoised image is then reconstructed as $\hat x = y - \hat r$.

The model is trained on paired LDCT/NDCT slices from the 2016 NIH--AAPM--Mayo Clinic Low-Dose CT Grand Challenge dataset\footnote{\url{https://www.aapm.org/grandchallenge/lowdosect/}}. Specifically, we use the subset reconstructed with the B30 kernel and a slice thickness of 1\,mm, which contains 5,926 paired NDCT--LDCT slices from 10 patients. Data from 9 patients are used for training, while all 526 paired slices from patient L506 are reserved for testing. \textit{Additional datasets will be evaluated in future versions of this work.}
Training is performed for 50{,}000 iterations using AdamW with a learning rate of $10^{-4}$. We adopt step decay (step 10{,}000, factor 0.5), exponential moving average (EMA) with decay 0.999, and gradient clipping with a maximum norm of 1.0. The mini-batch size is 24.

To accommodate different application needs, we further provide three model variants, namely RDDM-Fine, RDDM-Balanced, and RDDM-Smooth, corresponding to different trade-offs between detail preservation and noise suppression. Their parameter configurations are summarized in the bottom-right panel of Fig.~\ref{fig:overview}.

\begin{figure}[t]
    \centering
    \includegraphics[width=\linewidth]{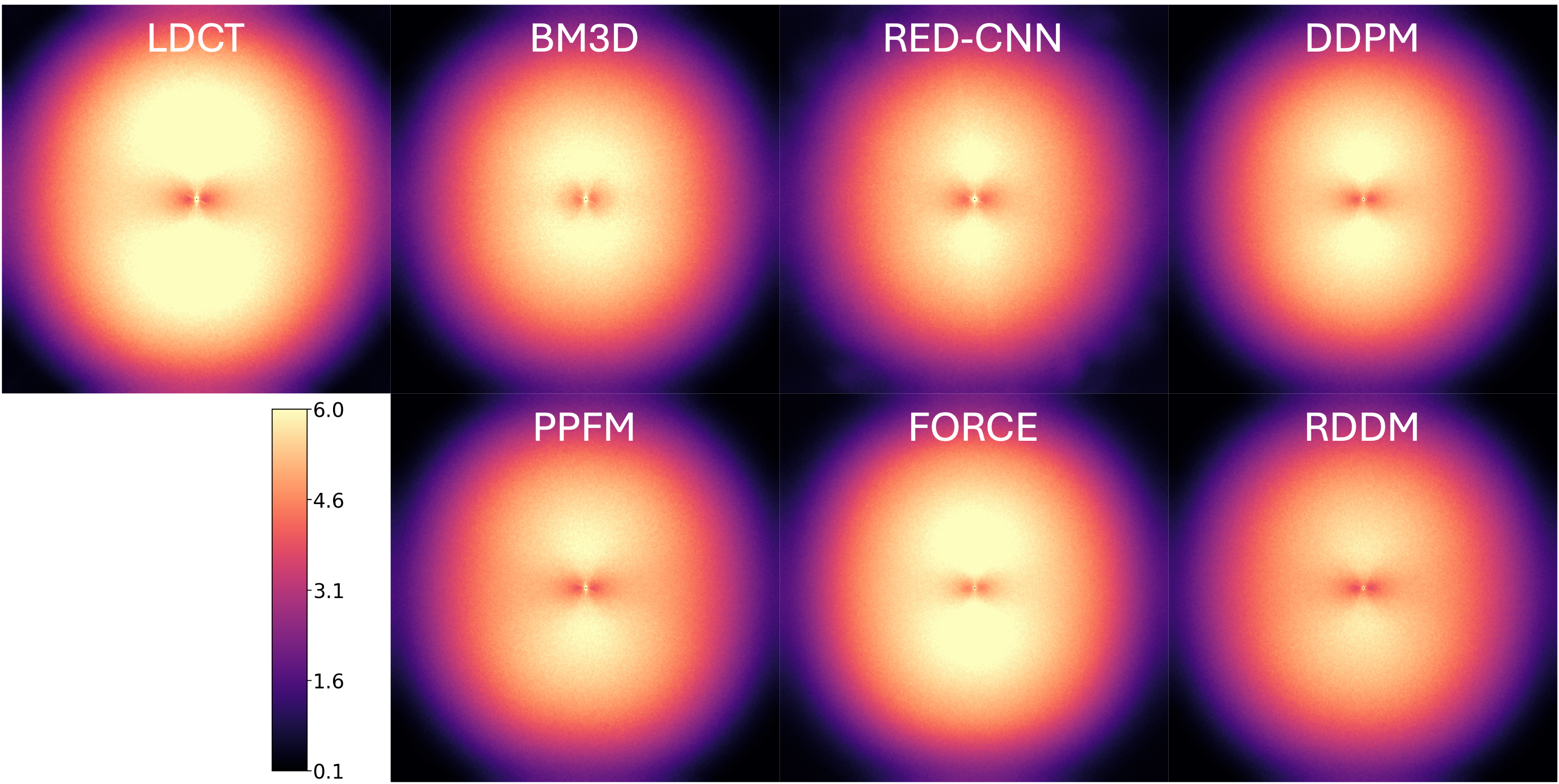}
    \caption{
    Residual power spectrum (RPS) visualization of LDCT and different denoising methods with respect to the NDCT reference on the quarter-dose Mayo test set. Here, RDDM refers to RDDM-Fine.
    }
    \label{fig:rps}
\end{figure}

\section{Experiments}
We conducted both qualitative and quantitative comparisons with classical and representative supervised denoising methods, including BM3D~\cite{dabov2007image}, RED-CNN~\cite{chen2017low}, CTformer~\cite{wang2023ctformer}, DDPM~\cite{ho2020denoising}, DDIM~\cite{song2020denoising}, PPFM~\cite{hein2024ppfm}, the supervised version of FORCE~\cite{xia2026tomographic}, and the original image-space Drifting Models~\cite{deng2026drifting}. 
Here, DDPM refers to a conditional DDPM trained on paired NDCT--LDCT images, while DDIM provides an accelerated inference scheme based on the same conditional DDPM framework. For FORCE, we evaluated both the default version with 500 NFEs and a fast version with only 50 NFEs. For Drifting Models, we used the default temperature setting reported in the original paper, $\mathcal{T}=\{0.02,0.05,0.2\}$.

To ensure a fair comparison, all learning-based methods were retrained on the same dataset and evaluated after convergence. Quantitative evaluation was performed using peak signal-to-noise ratio (PSNR), structural similarity index measure (SSIM)~\cite{wang2004image}, and Fréchet inception distance (FID). All experiments were conducted on a single NVIDIA H100 GPU with 80\,GB HBM3 memory.

\begin{figure*}[t]
    \centering
    \includegraphics[width=\textwidth]{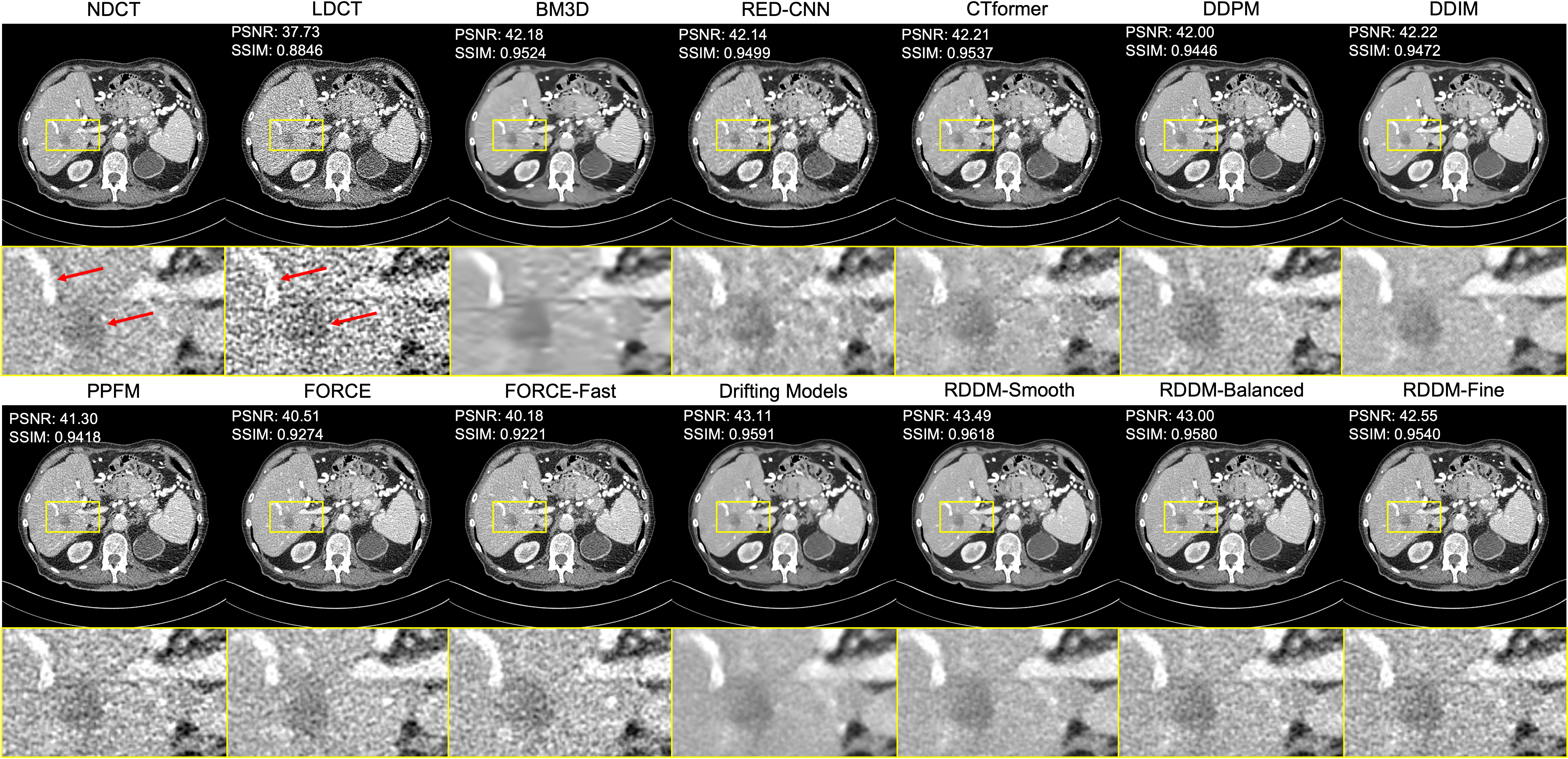}
    \caption{
    Visual comparison of denoising results from different methods on a quarter-dose sample from the Mayo dataset.
    The display window is $[-160,\,240]$ in Hounsfield unit (HU).
    }
    \label{fig:visual_comparison}
\end{figure*}

\begin{table*}[t]
\centering
\caption{Quantitative comparison of different methods on the quarter-dose Mayo test set. PSNR, SSIM, and inference time are reported as mean $\pm$ standard deviation over 526 test slices. Best and second-best results are highlighted in bold and underlined, respectively.}
\label{tab:quantitative_comparison}
\renewcommand{\arraystretch}{1.2}
\setlength{\tabcolsep}{16pt}

\begin{tabular}{lcccc}
\toprule
\textbf{Method} & \textbf{PSNR} $\uparrow$ & \textbf{SSIM} $\uparrow$ & \textbf{FID} $\downarrow$ & \textbf{Avg. inference time (s)} $\downarrow$ \\
\midrule

LDCT 
& 42.35 $\pm$ 1.79
& 0.9498 $\pm$ 0.0228
& 56.39
& -- \\

BM3D~\cite{dabov2007image} 
& 45.14 $\pm$ 1.12 
& 0.9736 $\pm$ 0.0071
& 84.92
& 1.49 $\pm$ 0.06 \\

RED-CNN~\cite{chen2017low} 
& 45.83 $\pm$ 1.37
& 0.9761 $\pm$ 0.0088
& 35.59
& \textbf{0.01} $\pm$ 0.01 \\

CTformer~\cite{wang2023ctformer} 
& 45.99 $\pm$ 1.41
& 0.9784 $\pm$ 0.0082
& 34.63
& 0.06 $\pm$ 0.01 \\

DDPM~\cite{ho2020denoising} 
& 46.10 $\pm$ 1.62
& 0.9755 $\pm$ 0.0106
& 12.46
& 25.22 $\pm$ 0.03 \\

DDIM~\cite{song2020denoising} 
& 45.68 $\pm$ 1.94
& 0.9721 $\pm$ 0.0155
& 14.94
& 1.25 $\pm$ 0.02 \\

PPFM~\cite{hein2024ppfm} 
& 45.74 $\pm$ 2.84
& 0.9757 $\pm$ 0.0113
& 8.37
& 0.07 $\pm$ 0.04 \\

FORCE~\cite{xia2026tomographic} 
& 44.08 $\pm$ 1.43
& 0.9651 $\pm$ 0.0134
& \underline{6.56}
& 18.93 $\pm$ 0.03 \\

FORCE-Fast~\cite{xia2026tomographic}
& 44.02 $\pm$ 1.45
& 0.9645 $\pm$ 0.0138
& 6.66
& 1.87 $\pm$ 0.03 \\

Drifting Models~\cite{deng2026drifting}
& 45.94 $\pm$ 1.32
& \underline{0.9806} $\pm$ 0.0075
& 37.92
& \underline{0.02} $\pm$ 0.02 \\

\midrule
RDDM-Smooth 
& \textbf{47.33} $\pm$ 1.47
& \textbf{0.9828} $\pm$ 0.0071
& 32.09
& \underline{0.02} $\pm$ 0.01 \\

RDDM-Balanced 
& \underline{46.79} $\pm$ 1.47
& \underline{0.9808} $\pm$ 0.0078
& 9.13
& \underline{0.02} $\pm$ 0.01 \\

RDDM-Fine 
& 46.29 $\pm$ 1.49
& 0.9786 $\pm$ 0.0088
& \textbf{5.87}
& \underline{0.02} $\pm$ 0.01 \\

\bottomrule
\end{tabular}
\end{table*}

\subsection{Qualitative Study}
For visual comparison, Fig.~\ref{fig:visual_comparison} shows representative slices from the Mayo test set together with the denoising results of all methods. This example corresponds to an abdominal CT image with a liver metastasis. The regions of interest (ROIs) containing the metastasis are highlighted by yellow boxes, while the lesion and a representative anatomical structure are indicated by red arrows.

In this comparison, BM3D yielded an overly smooth denoised image, with noticeably distorted and blurred textures across different tissues. RED-CNN and CTformer preserved the major anatomical structures, but fine textures remained blurred and noticeable residual artifacts were still present.
Among the diffusion- and flow-based baselines, DDPM achieved a relatively good balance between noise suppression and detail preservation. Nevertheless, its reconstruction still showed noticeable inconsistencies with the NDCT reference in the tissue region indicated in the upper-left corner, where the tissue continuity observed in NDCT was disrupted. Compared with DDPM, DDIM improved inference efficiency, but this gain came at the cost of smoother reconstructions with fewer fine details, which may increase the risk of blurring clinically important anatomical structures. PPFM pursued higher fidelity to NDCT, but at the cost of excessive residual noise. FORCE showed higher visual consistency with NDCT than PPFM or DDPM. However, similar to DDPM, it still exhibited noticeable discrepancies from NDCT in multiple regions. This observation is consistent with its lowest pixel-level evaluation scores (PSNR and SSIM), suggesting a higher risk of altering true anatomical and textural information. Compared with FORCE, FORCE-Fast reduced the inference cost, but at the expense of more residual noise and damaged structural details. The original Drifting Models suppressed noise too aggressively, preserving only the major anatomical framework while losing important intra-tissue texture details.

Among the three RDDM variants, RDDM-Smooth achieved the highest PSNR and SSIM among all methods, while still preserving clear tissue structures without becoming excessively smooth. These results make it a practical solution when strong noise suppression is desired without sacrificing overall reconstruction fidelity. In contrast, RDDM-Fine generated a denoised result that was highly consistent with NDCT, preserving nearly all texture details, including the underlying noise characteristics. Such high-fidelity reconstructions are often more desirable in clinical diagnosis. RDDM-Balanced provided a compromise between smoothness and detail preservation, maintaining the anatomical structures and tissue textures present in NDCT while achieving a lower noise level than NDCT. Overall, RDDM offered a range of favorable trade-offs between noise suppression and visual fidelity, making it well suited for diverse clinical needs.

We further analyzed the residual power spectrum (RPS) of several representative methods to better characterize the frequency-domain error relative to NDCT. The RPS describes the spectral energy of the residual between a test image and its NDCT reference. Fig.~\ref{fig:rps} shows the RPS visualizations of LDCT and all denoised results, aggregated over the entire Mayo test set.
The RPS of LDCT exhibits a bright central region spreading over a wide frequency range, together with clear vertical patterns, indicating that the original residual is dominated by structured noise. The RPS of BM3D shows strong and uniformly distributed energy in the low-frequency region, suggesting over-smoothed reconstructions and loss of texture information. 
RED-CNN, DDPM, and PPFM suppress noise over a broad frequency range; however, their RPS maps still contain bright vertical streaks, indicating residual structured patterns and insufficient suppression of edge-related structures. Among these methods, PPFM exhibits relatively better overall performance, with weaker residual streaks in the RPS.
FORCE exhibits very prominent vertical streaks and excessively strong, broadly distributed low-frequency energy in the RPS, implying lower pixel-level consistency between its reconstruction and the NDCT reference.
In contrast, RDDM achieves the most effective suppression of low-frequency residual energy among all compared methods. Its RPS exhibits the smallest and darkest circular pattern, indicating the lowest residual magnitude across frequencies. This suggests that its reconstruction is highly consistent with NDCT at the pixel level.

\subsection{Quantitative Study}
We quantitatively compared the denoising performance of different methods on the held-out Mayo test set. Table~\ref{tab:quantitative_comparison} summarizes the results of all methods.
All variants of RDDM achieved higher PSNR and SSIM than the competing methods, including the original Drifting Models, among which RDDM-Smooth obtained the highest PSNR and SSIM. RDDM-Fine achieved the best FID of 5.87 on the test set, outperforming FORCE (6.56). These results indicate that RDDM achieves superior consistency with NDCT at both the pixel level and the distribution level, demonstrating strong capability in noise suppression and structure preservation.
In terms of inference efficiency, RDDM retains the one-step inference advantage of Drifting Models and requires only 1 NFE during denoising, making it substantially more efficient than conventional diffusion- and flow-based methods. The inference time of RDDM for a single Mayo slice is only about 15 ms, comparable to RED-CNN and among the fastest of all methods.
Overall, RDDM achieves a strong balance between denoising accuracy and computational efficiency, and its high-fidelity reconstructions are well suited for clinical diagnosis and downstream analysis tasks.

\begin{figure}[t]
    \centering
    \includegraphics[width=\linewidth]{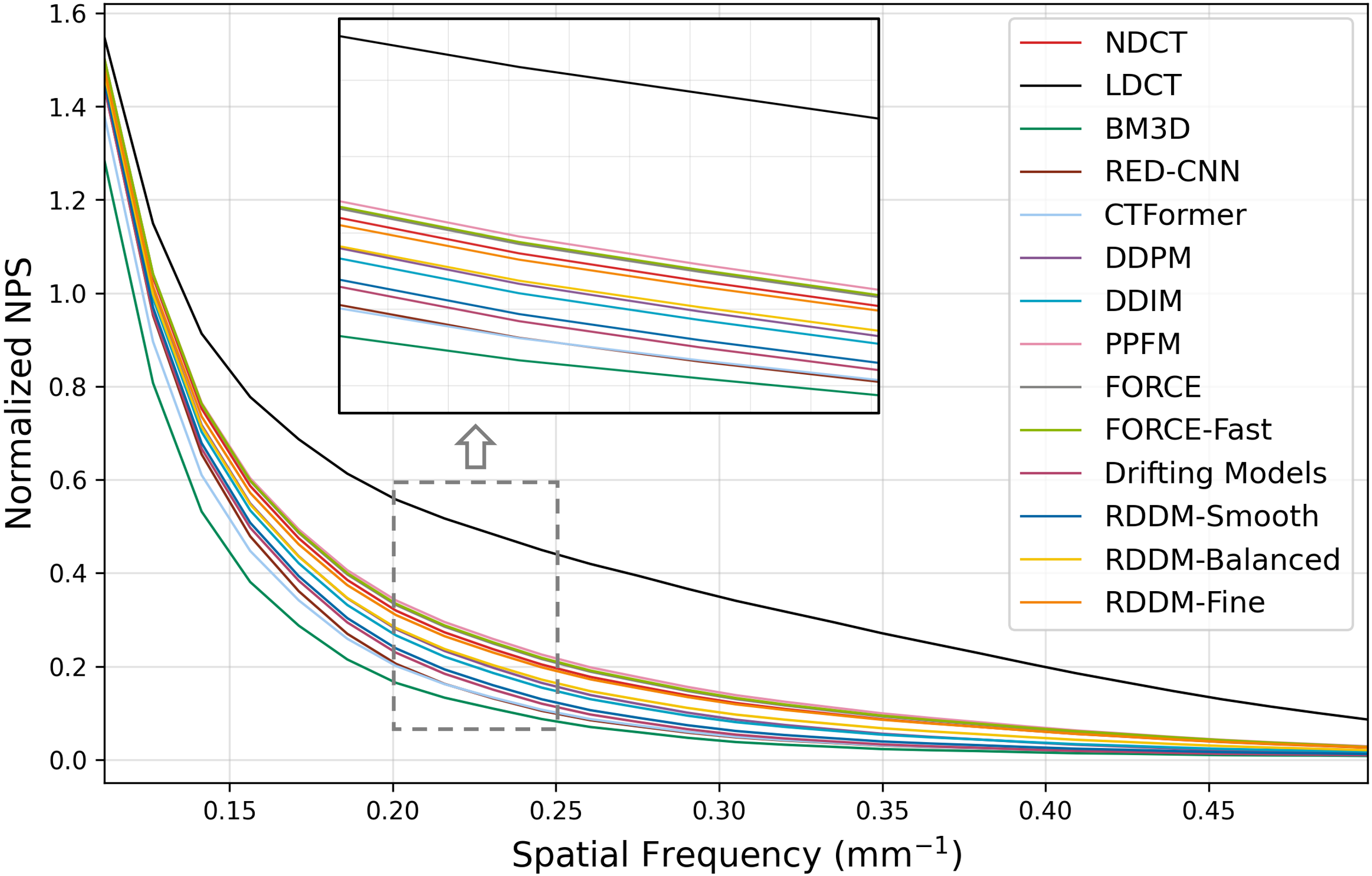}
    \caption{Normalized radial 1D noise power spectrum (NPS) profiles averaged over the quarter-dose Mayo test set.}
    \label{fig:nps}
\end{figure}

\begin{table*}[t]
\centering
\caption{Analysis of the temperature parameter $\tau$ in the residual drifting field on the quarter-dose Mayo test set. PSNR and SSIM are reported as mean $\pm$ standard deviation over 526 test slices. Best results are highlighted in bold.}
\label{tab:temperature_ablation}
\renewcommand{\arraystretch}{1.2}
\setlength{\tabcolsep}{10pt}

\begin{tabular}{lccc lccc}
\toprule

\textbf{Single-$\tau$} & \textbf{PSNR} $\uparrow$ & \textbf{SSIM} $\uparrow$ & \textbf{FID} $\downarrow$ 
& \textbf{Multi-$\tau$} & \textbf{PSNR} $\uparrow$ & \textbf{SSIM} $\uparrow$ & \textbf{FID} $\downarrow$ \\
\cmidrule(lr){1-4} \cmidrule(lr){5-8}

$\tau=0.02$ & 42.35 $\pm$ 1.79 & 0.9498 $\pm$ 0.02 & 56.39 
& $\mathcal{T}=\{0.02,0.05,0.2\}$ & 46.95 $\pm$ 1.47 & 0.9815 $\pm$ 0.01 & 11.97 \\

$\tau=0.05$ & 42.38 $\pm$ 1.79 & 0.9498 $\pm$ 0.02 & 56.19 
& $\mathcal{T}=\{0.05,0.2\}$ & 46.94 $\pm$ 1.46 & 0.9815 $\pm$ 0.01 & 11.70 \\

$\tau=0.1$  & \textbf{47.24} $\pm$ 1.45 & \textbf{0.9826} $\pm$ 0.01 & 43.36 
& $\mathcal{T}=\{0.1,1.0\}$ & 46.81 $\pm$ 1.43 & 0.9810 $\pm$ 0.01 & 9.93 \\

$\tau=0.2$  & 46.95 $\pm$ 1.47 & 0.9815 $\pm$ 0.01 & 12.08
& $\mathcal{T}=\{0.1,1.5\}$ & \textbf{47.11} $\pm$ 1.44 & \textbf{0.9822} $\pm$ 0.01 & 18.13 \\

$\tau=0.5$  & 45.06 $\pm$ 1.52 & 0.9720 $\pm$ 0.01 & 13.65 
& $\mathcal{T}=\{0.2,1.0\}$ & 46.79 $\pm$ 1.47 & 0.9808 $\pm$ 0.01 & 9.13 \\

$\tau=1.0$  & 46.04 $\pm$ 1.49 & 0.9774 $\pm$ 0.01 & \textbf{6.15} 
& $\mathcal{T}=\{0.2,1.5\}$ & 46.90$\pm$ 1.47 & 0.9815 $\pm$ 0.01 & 10.77 \\

$\tau=1.5$  & 46.58 $\pm$ 1.49 & 0.9799 $\pm$ 0.01 & 7.25 
& $\mathcal{T}=\{0.2,2.0\}$ & 46.95 $\pm$ 1.47 & 0.9815 $\pm$ 0.01 & 11.65 \\

$\tau=2.0$  & 46.83 $\pm$ 1.48 & 0.9810 $\pm$ 0.01 & 10.51 
& $\mathcal{T}=\{1.0,1.5\}$ & 46.29 $\pm$ 1.49 & 0.9786 $\pm$ 0.01 & \textbf{5.87} \\

$\tau=3.0$  & 47.06 $\pm$ 1.46 & 0.9820 $\pm$ 0.01 & 16.50
& $\mathcal{T}=\{1.0,2.0\}$ & 46.37 $\pm$ 1.48 & 0.9789 $\pm$ 0.01 & 6.06 \\

\bottomrule
\end{tabular}
\end{table*}

\begin{table*}[t]
\centering
\caption{Analysis of the $\ell_1$ loss weight $\lambda$ under single- and multi-temperature settings on the quarter-dose Mayo test set. PSNR and SSIM are reported as mean $\pm$ standard deviation over 526 test slices.}
\label{tab:l1_ablation}
\renewcommand{\arraystretch}{1.2}
\setlength{\tabcolsep}{6pt}

\begin{tabular}{lccccccc}
\toprule
& \multicolumn{3}{c}{\textbf{Single-$\tau$} ($\tau=1.0$)} 
& \multicolumn{3}{c}{\textbf{Multi-$\tau$} ($\mathcal{T}=\{1.0,2.0\}$)} 
& \multirow{2}{*}{\textbf{$\ell_1$-only}} \\
\cmidrule(lr){2-4} \cmidrule(lr){5-7}

\textbf{Metric} 
& $\lambda=0$ & $\lambda=0.001$ & $\lambda=0.01$
& $\lambda=0$ & $\lambda=0.001$ & $\lambda=0.01$
& \\
\midrule

PSNR $\uparrow$
& 46.04 $\pm$ 1.49
& 46.74 $\pm$ 1.57
& 47.33 $\pm$ 1.47
& 46.37 $\pm$ 1.48
& 46.79 $\pm$ 1.54
& 47.31 $\pm$ 1.47
& 47.31 $\pm$ 1.48 \\

SSIM $\uparrow$
& 0.9774 $\pm$ 0.01
& 0.9805 $\pm$ 0.01
& 0.9828 $\pm$ 0.01
& 0.9789 $\pm$ 0.01
& 0.9807 $\pm$ 0.01
& 0.9828 $\pm$ 0.01
& 0.9828 $\pm$ 0.01 \\

FID $\downarrow$
& 6.15
& 8.88
& 32.09
& 6.06
& 9.32
& 32.09
& 43.10 \\

\bottomrule
\end{tabular}
\end{table*}

To further characterize the noise properties of the denoised results, we analyzed the noise power spectrum (NPS) for all methods. Specifically, four flat ROIs of size $64\times64$ were selected from each of the 526 test slices. These ROIs were carefully chosen to avoid structured anatomical content, such as tissues or textures, thereby minimizing contamination of the NPS by structural information. Fig.~\ref{fig:nps} shows the normalized radial 1D NPS curves aggregated over the entire Mayo test set. The curves are displayed within the most discriminative frequency range of $0.1$--$0.5~\mathrm{mm}^{-1}$, with a further zoom-in on $0.2$--$0.25~\mathrm{mm}^{-1}$ for clearer comparison.

The NPS curves of PPFM, FORCE, FORCE-Fast, and RDDM-Fine are all close to that of NDCT, indicating that the noise characteristics of their reconstructions are highly consistent with those of NDCT. Among them, the curves of PPFM, FORCE, and FORCE-Fast lie above the NDCT curve, suggesting excessive residual noise in the denoised results. In contrast, the curve of RDDM-Fine lies slightly below and closely follows the NDCT curve. Since NDCT images are acquired at clinically acceptable radiation doses, they still contain a certain level of noise. From this perspective, a desirable NPS curve after denoising should lie slightly below that of NDCT, reflecting effective noise suppression without excessive residual noise or over-smoothing. Accordingly, RDDM-Fine achieves the most appropriate level of noise reduction among all compared methods. This observation is also consistent with the visual comparisons in Fig.~\ref{fig:visual_comparison}, where RDDM-Fine effectively suppresses noise while preserving structural details, whereas FORCE and PPFM tend to retain noticeable noise.
DDPM shows a similar NPS trend to RDDM-Balanced, but requires substantially longer denoising time due to its multi-step inference process. Although DDIM reduces the number of inference steps compared with DDPM, its NPS curve deviates more from that of NDCT, indicating a trade-off between inference efficiency and denoising performance. In contrast, BM3D, RED-CNN, CTformer, and RDDM-Smooth exhibit NPS curves far below the desired region, indicating aggressive noise suppression and correspondingly smoother denoised results. In particular, BM3D shows the lowest noise power across all methods, which is consistent with its visibly over-smoothed appearance.
In clinical practice, such overly smooth reconstructions may obscure subtle anatomical structures and attenuate diagnostically relevant tissue textures, thereby compromising diagnostic confidence and subsequent clinical interpretation.

Considering NPS characteristics together with visual quality and computational efficiency, RDDM-Fine demonstrates the best overall performance and shows strong potential for practical clinical applications. Meanwhile, the other two RDDM variants also remain valuable for scenarios requiring different levels of noise suppression.

\subsection{Temperature Analysis in the Residual Drifting Field}
We analyzed the temperature parameter $\tau$ in the similarity kernel of the residual drifting field, as defined in~\eqref{eq:similarity_kernel}. Specifically, we evaluated both single-temperature settings ranging from $0.02$ to $3.0$ and several multi-temperature configurations. The quantitative results on the Mayo test set are summarized in Table~\ref{tab:temperature_ablation}.

The temperature parameter directly affects the shape of the similarity kernel and thus the strength and range of the attractive and repulsive interactions in the drifting field. When the temperature is too low, the similarity kernel becomes highly peaked, causing the attractive and repulsive forces to fluctuate rapidly and making the model behavior unstable. 
As shown in Table~\ref{tab:temperature_ablation}, very low single-temperature settings, such as $\tau=0.02$ and $\tau=0.05$, lead to poor overall denoising quality and larger FID values, indicating a stronger deviation from the NDCT distribution.
In contrast, when the temperature is too high, the similarity kernel becomes overly flat. In this case, the attractive and repulsive forces become too weak, resulting in slow convergence and reduced effectiveness of the drifting process. This also leads to suboptimal denoising performance and increased FID.

Compared with single-temperature settings, the multi-temperature formulation provides more stable denoising performance and is less sensitive to extreme temperature choices. By combining local interactions induced by low temperatures with more non-local interactions introduced by higher temperatures, it achieves a better balance between structural detail preservation and noise suppression. Notably, $\mathcal{T}=\{0.02,0.05,0.2\}$ is the default temperature setting recommended in Drifting Models~\cite{deng2026drifting}, enabling a direct comparison with the corresponding Drifting Models result in Table~\ref{tab:quantitative_comparison}. Under this setting, RDDM still achieves substantially better denoising performance, further highlighting the advantage of the RDDM framework for LDCT denoising.

\subsection{Effect of Pixel-Level Supervision}
We analyzed the effect of incorporating the $\ell_1$ reconstruction loss into the RDDM objective. Specifically, we evaluated $\lambda \in \{0,\,0.001,\,0.01\}$ under both a single-temperature setting ($\tau=1.0$) and a multi-temperature setting ($\mathcal{T}=\{1.0,2.0\}$), together with an $\ell_1$-only baseline. The quantitative results are summarized in Table~\ref{tab:l1_ablation}.

Under both temperature settings, introducing the $\ell_1$ loss improves PSNR and SSIM, indicating better pixel-level consistency with NDCT and stronger noise suppression. However, this stronger suppression also lowers the residual noise level below that of NDCT and leads to degraded FID. When the $\ell_1$ loss weight reaches $0.01$, the final PSNR and SSIM approach those of the $\ell_1$-only solution, indicating that the $\ell_1$ term gradually dominates the optimization in the later stages. Nevertheless, compared with the $\ell_1$-only baseline, RDDM with $\lambda = 0.01$ still achieves clearly better FID while maintaining comparable PSNR and SSIM. This suggests that, under strong pixel-level supervision, RDDM can still provide stronger distributional consistency than conventional $\ell_1$-only training, while achieving strong noise suppression and good structural preservation.

We also observed that both changing the temperature and introducing the $\ell_1$ loss can improve pixel-level reconstruction quality. For example, taking the setting $(\tau=1.0,\lambda=0)$ in Table~\ref{tab:l1_ablation} as the reference, either lowering the temperature to $\tau=0.1$ in Table~\ref{tab:temperature_ablation} or introducing $\lambda=0.01$ in Table~\ref{tab:l1_ablation} improves PSNR and SSIM, although both lead to some loss in FID. Compared with lowering the temperature alone, adding the $\ell_1$ loss yields higher PSNR and SSIM with a smaller degradation in FID. This indicates that combining the drifting loss with pixel-level supervision provides a more effective way to achieve strong noise suppression while maintaining favorable reconstruction fidelity.

\section{Conclusion}
In this work, we proposed RDDM, a low-dose CT denoising method based on Drifting Models and residual learning. RDDM treats the residual in LDCT as the attractive force and the reconstructed residual generated by the model as the repulsive force. Based on these two forces, a residual drifting field is constructed. During training, this drifting field progressively evolves an initial random noise distribution toward the latent residual distribution in LDCT. By absorbing the multi-step distribution evolution into model training, RDDM requires only one-step inference to achieve efficient and effective denoising. Experimental results demonstrate that RDDM not only provides strong noise suppression, but also preserves important anatomical structures and fine texture details. Its reconstructions are highly consistent with NDCT at both the pixel level and the distribution level, delivering superior image quality and fidelity compared with state-of-the-art methods. These results highlight the strong potential of RDDM for practical clinical LDCT denoising and related downstream analysis tasks.

\section*{Acknowledgments}
This project is funded by the National Institute of Health R01 CA233888. The content is solely the responsibility of
the authors and does not necessarily represent the official views of the National Institutes of Health. 

\bibliographystyle{IEEEtran}
\bibliography{reference}  
\vfill

\end{document}